\documentclass[10pt,aps,onecolumn,prl,showpacs,notitlepage]{revtex4-1}
\usepackage{graphicx}
\usepackage{color}
\usepackage{amsmath, amsthm, amssymb}
\usepackage[bf,Large]{}
\begin{document}
\title{Fermi-surface-free superconductivity in underdoped (Bi,Pb)$_{2}$(Sr,La)$_{2}$CuO$_{6+\delta}$ (Bi2201)} 
\author{Peter Mistark*, Hasnain Hafiz, Robert S. Markiewicz, Arun Bansil}
\affiliation{Department of Physics, Northeastern University, Boston}
\date{Version of \today}
\begin{abstract}
We show that hole-doped cuprates can harbor {\it Fermi-surface-free superconductivity} similar to the case of the pnictides. This occurs near the doping at which a new Fermi surface pocket appears in the antinodal region. The change in Fermi surface topology is accompanied by a characteristic rise in spectral weight. Our results support the presence of a trisected superconducting dome, and suggest that superconductivity is responsible for stabilizing the $(\pi,\pi)$ magnetic order.
\end{abstract}
\maketitle

Recent ARPES studies on pnictides adduce that the superconducting transition temperature $T_c$ depends sensitively on details of the band structure and Fermi surface (FS)\textsuperscript{1}. In particular, proximity of the FS to a band edge and the associated Van Hove singularity (VHS) correlates with significantly enhanced $T_c$'s. An investigation of Ba$_{1-x}$K$_x$Fe$_2$As$_2$\textsuperscript{2} indicates that when a band edge approaches the FS, superconductivity can be observed even before the band crosses the FS. Bang\textsuperscript{2} suggests that this FS-free superconductivity is driven by the shadow bands resulting from the symmetrization of spectral weight around the Fermi energy and the formation of a related ‘shadow gap’ in the BCS theory. We show how this remarkable effect can also arise in the single-band case of the cuprates when the band is split by magnetic order. 

In the electron doped system it is well known that the $(\pi,\pi)$ antiferromagnetic (AF) order can induce two distinct {\it ‘topological transitions’} (TT's) with doping\textsuperscript{3,4}. At half-filling, the AF order splits the band into upper and lower magnetic bands (U/LMBs), and low electron doping moves the FS into the bottom of the UMB.  As the electron doping increases, the LMB moves up in energy and eventually crosses the FS leading to the emergence of hole pockets around $(\pi/2,\pi/2)$. This is the first topological transition (TT1) in this system. The second topological transition (TT2) occurs at higher doping when the electron and hole pockets merge into the single continuous FS of the paramagnetic system. 

In this paper we show that a transition similar to TT1 can occur in hole doped cuprates such as (Bi,Pb)$_{2}$(Sr,La)$_{2}$CuO$_{6+\delta}$  (Bi2201)\textsuperscript{5}. This transition is however different in that the first holes now enter the LMB, so the transition occurs when the UMB moves down in energy and crosses the FS, introducing electron pockets around $(\pi,0)$\textsuperscript{6,7}. It is near this transition that we find evidence for FS-free d-wave superconductivity. 

\section{Results}

\begin{figure}[htp] 
\includegraphics[scale=0.9,keepaspectratio=true]{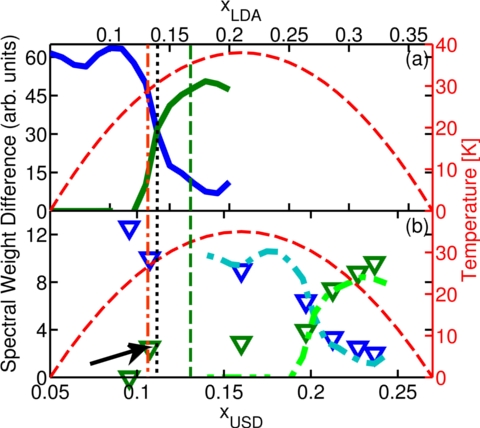}
\caption
{ 
{\bf Theoretical and experimental pseudogap spectral weight and AN pair weight.} Theoretical (a) and experimental\textsuperscript{5} (b) pseudogap spectral weight (blue line and markers) and AN pair weight in Bi2201 (green lines and markers). The red dashed curve shows the superconducting dome, $T_c(x)$, with temperature on the right hand vertical axis. The values for $T_c$ are estimated as $\Delta_{SC}=5k_BT_c$\textsuperscript{14} and the SC dome is assumed parabolic and given by $\Delta_0=0.01637[1-39.0625(0.21-x_{LDA})^2]$. Vertical lines spanning (a) and (b) represent the beginning of TT1 as determined in this work (black dotted), thermopower\textsuperscript{20} (green dashed), and STM\textsuperscript{15} (orange dot-dashed) experiments. The black arrow in (b) points to the onset of AN weight in experimental data. Light blue and green dot-dashed lines in (b) represent our data in (a) shifted by $x_{USD}=0.0903$ and scaled by $5/30$. The similarity of the second AN weight increase in the data at higher doping to TT1 suggests an additional TT probably associated with a competing CDW order.
}
\label{fig:1}
\end{figure}

Our analysis is based on a one band mean-field Hubbard model with competing $(\pi,\pi)$-antiferromagnetic ($(\pi,\pi)$-AF) and d-wave superconducting (dSC) orders, which we have invoked previously in connection with electron-doped cuprates\textsuperscript{8}. Using quasi-particle GW (QPGW)\textsuperscript{9} self-energy corrections, we have shown that this model provides a reasonable description of many salient features of the electronic spectra of the cuprates as observed in ARPES\textsuperscript{10} and other spectroscopies\textsuperscript{11,12}. The Hamiltonian is
\begin{eqnarray}\label{HFHam}
H&=&\sum_{{\bf k},\sigma}(\epsilon_{{\bf k}}-\epsilon_F)c^{\dag}_{{\bf k},\sigma}
c_{{\bf k},\sigma}+\Delta_{AF}\sum_{{\bf k},{\bf k}^{\prime}}[c^{\dag}_{{\bf k}+{\bf Q},\uparrow}
c_{{\bf k},\uparrow}-c^{\dag}_{{\bf k}^{\prime}-{\bf Q},\downarrow}
c_{{\bf k}^{\prime},\downarrow}]\sum_{{\bf k}}[\Delta_{\bf k}c^{\dag}_{{\bf k},\uparrow}
c^{\dag}_{-{\bf k},\downarrow}+\Delta^*_{\bf k}c_{-{\bf k},\downarrow}c_{{\bf k},\uparrow}],
\end{eqnarray}
where $\epsilon_F$ is the Fermi energy, $c^{\dag}_{{\bf k},\sigma}$ and
$c_{{\bf k},\sigma}$ are the creation and annihilation operators for an electron of momentum ${\bf k}$ and spin $\sigma$, respectively. $\Delta_{AF}=US$ is the AF gap with $U$ the Hubbard U and $S$ the staggered magnetization. $\Delta_{\bf k}=\Delta_0(cos(k_xa)-cos(k_ya))/2$ is the d-wave superconducting gap, where $\Delta_0$ is the maximum gap, and $\epsilon_{\bf k}$  gives the bare band dispersion as 
\begin{eqnarray}\label{tbBand}
\epsilon_{\bf k}&=&-2t[c_x(a)+c_y(a)]\nonumber\\
&&-4t^{\prime}c_x(a)c_y(a)-2t^{\prime\prime}[c_x(2a)+c_y(2a)] \nonumber\\
&&-4t^{\prime\prime\prime}[c_x(2a)c_y(a)+c_y(2a)c_x(a)]\nonumber\\
&&-4t^{iv}c_x(2a)c_y(2a),
\end{eqnarray}
where $t^i$ are the hopping parameters, $c_i(\alpha a)=cos(\alpha {\bf k}_ia)$, and $a$ is the lattice constant.  The values of the $t$'s are taken from photoemission experiment\textsuperscript{13}. 
The AF gap in Eq.(1) splits the quasiparticle spectrum into upper ($\nu=+$) and lower ($\nu=-$) magnetic bands (U/LMB), which are further split by superconductivity, yielding dispersions
\begin{eqnarray}\label{eigen}
E_{\bf k}^{\nu}=\pm\sqrt{\left(E^{s,\nu}_{k}\right)^2+\Delta_{k}^2}.
\end{eqnarray}
Here, $\xi_{\bf k}=\epsilon_{\bf k}-\epsilon_F$  and $\xi_{\bf k}^{\pm}=(\xi_{\bf k}\pm\xi_{{\bf k}+{\bf Q}})/2$.  $E_{\bf k}^{s,\nu}=\xi_{\bf k}^++\nu E_{0k}$ describes the quasiparticles in the non-superconducting state with AF order, and $E_{0{\bf k}}=\sqrt{\left(\xi_{\bf k}^-\right)^2+(US)^2}$. The diagonalization results in coherence factors for the two bands as follows:  
\begin{eqnarray}\label{eigenvec}
\alpha_{\bf k}(\beta_{\bf k})&=&\sqrt{(1\pm\xi_{\bf k}^-/E_{0{\bf k}})/2},\nonumber\\
u_{\bf k}^{\nu} (v_{\bf k}^{\nu})&=&\sqrt{[1\pm(\xi^+_{\bf k}+\nu E_{0{\bf k}})/E^{\nu}_{\bf k}]/2}.
\end{eqnarray}
We can write equations for the gap $\Delta_0$ and the staggered magnetization, $S$, in terms of the coherence factors as
\begin{eqnarray}
\Delta_{0}&=&-V\sum_{{\bf k}}g_{{\bf k}}\left[u_{{\bf k}}^{+}v_{{\bf k}}^{+}\tanh{}(\beta E_{{\bf k}}^{+}/2)+u_{{\bf k}}^{-}v_{{\bf k}}^{-}\tanh{}(\beta E_{{\bf k}}^{-}/2)\right]\nonumber\\
&=&-V\Delta_0\sum_{{\bf k}}g_{{\bf k}}^2 \left[\frac{1}{2E_{{\bf k}}^{+}}\tanh{}(\beta E_{{\bf k}}^{+}/2)+\frac{1}{2E_{{\bf k}}^{-}}\tanh{}(\beta E_{{\bf k}}^{-}/2)\right],
\end{eqnarray}
\begin{eqnarray}
S &=&\frac{1}{N}\sum_{{\bf k}}\alpha_{{\bf k}}\beta_{{\bf k}}\big[\big((v_{{\bf k}}^{-})^2-(v_{{\bf k}}^{+})^2\big)+\big((v_{{\bf k}}^{+})^2-(u_{{\bf k}}^{+})^2\big)f(E_{{\bf k}}^{+})-\big((v_{{\bf k}}^{-})^2-(u_{{\bf k}}^{-})^2\big)f(E_{{\bf k}}^{-})\big],\nonumber\\
&=&\frac{US}{N}\sum_{{\bf k}}\frac{1}{4E_{0{\bf k}}} \left[\frac{E^{s,+}_{\bf k}}{E_{\bf k}^+}\tanh{}(\beta E_{{\bf k}}^{+}/2)-\frac{E^{s,-}_{\bf k}}{E_{\bf k}^-}\tanh{}(\beta E_{{\bf k}}^{-}/2)\right],
\end{eqnarray}
where $g_{{\bf k}}=cos(k_x)-cos(k_y)$. Here we assume that $\Delta_0$ forms a parabolic dome in doping\textsuperscript{14} with maximum at $x=0.21$ based on fits of the Fermi energy to experiment\textsuperscript{15}. For our hopping parameters, the VHS in the AF+SC system is found around $x=0.37$. We take the superconducting dome to terminate at the VHS\textsuperscript{16}. This gives a SC dome which starts at $x=0.05$, peaks at $x=0.21$, and terminates at $x=0.37$. Equation (5) then determines $V$.  Dopings based on the LDA band structure will be labeled as $x_{LDA}$, see Fig.~\ref{fig:1}(a). However, experimental data are often described in terms of a `universal superconducting dome' (USD) with optimal $T_c$ at $x_{USD}=0.16$,\textsuperscript{17}. The transformation of $x_{USD}$ to $x_{LDA}$ is given by $x_{LDA}=(32/22)~x_{USD}-0.022727$. Fig.~\ref{fig:1} compares these two doping scales.

\begin{figure}[htp] 
\includegraphics[scale=0.9,keepaspectratio=true]{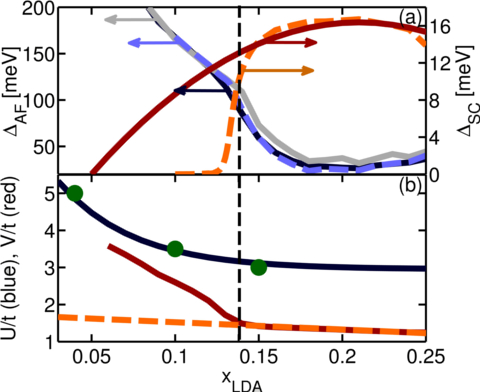}
\caption
{
{\bf Doping dependence of order parameters and corresponding potentials.} (a) Values of $\Delta_{AF}$ as a function of doping for a system with AFM order only (grey) or with combined SC+AF order (dark blue). The red curve shows the superconducting gap with the scale on the right hand vertical axis. The black dashed line indicates TT1 for our model at $x_{LDA}=0.138$. (b) $U/t$ fit (blue curve) to the results from Ref. 3 (green circles) as a function of doping and $V/t$ (red curve) calculated with equation (5) from the assumed superconducting dome. For the present analysis we are only interested in dopings greater than $x=0.05$, where the fit is quite good. The light red and light blue dashed curves in (a) and (b) represent the same quantities as their darker, solid counterparts except the doping dependence of $V$ is assumed linear and $\Delta_{SC}$ and $S$ are calculated using Eqs. 5 and 6. This shows that a large potential $V$ is needed for superconductivity to be sustained to dopings well below the first TT.
\label{fig:2}
}
\end{figure}

For a given doping $x$, we determine $\epsilon_F$ and $S$ self-consistently by using Luttinger's theorem to obtain $\epsilon_F$ from $x$, and $S$ from equation (6). The results are shown in Fig.~\ref{fig:2}. The Hubbard $U(x)$ is taken as a screened Coulomb potential, which has been studied extensively\textsuperscript{3,18,19}, Fig.~\ref{fig:2}(b). For this paper we used the data for the effective $U/t$ calculated in Ref.\textsuperscript{3}, fit to a decaying exponential $U/t=a_1e^{-x/x_0}+a_2$ where $x$ is the doping $a_1=4.6263$,  $a_2=2.95$, and $x_0= 0.045$. The first holes create a hole pocket in the LMB near the $(\pi/2,\pi/2)$ point in the Brillouin zone. The model has two transitions in the FS topology as a function of doping.  The magnetic gap $\Delta_{AF}$ decreases with doping, and goes to zero at some critical value of $x$ where the large nonmagnetic FS is restored (TT2). Before this happens, however, the UMB crosses the Fermi level, giving rise to an electron pocket near $\Gamma$. This is TT1 in Bi2201 in the vicinity of which the FS-free superconductivity arises. 

A recent ARPES study in underdoped Bi2201 determined the spectral weight loss associated with two transitions as a function of doping and temperature\textsuperscript{5}.  They found that the antinodal spectral weight began to decrease linearly with $T$ below a temperature scale $T^*(x)$, the pseudogap, and then the slope changed to a second value below a scale $T_{AN}(x)$, where $T^*>T_{AN}>T_c$, the superconducting temperature.  $T_{AN}$ signals the onset of superconducting pair fluctuations in the antinodal (AN) region near $(\pi,0)$, and it is determined at the point on the FS nearest to the antinode.  Here we analyze the $T=0$ limits of the weight loss, where the pseudogap weight loss is taken as the linear extrapolation of the weight loss above $T_{AN}$ to $T=0$, and the antinodal SC weight loss is the difference between this value and the total weight loss at $T=0$.  In our model, the pseudogap is assumed to be associated with $(\pi,\pi)$-AF order.  

The experimental data\textsuperscript{5} display a remarkable evolution of low-temperature spectral weight with doping $x$. While the pseudogap spectral weight decreases with doping (blue symbols in Fig.~\ref{fig:1}(b)), the AN pair weight, green symbols in Fig.~\ref{fig:1}(b), remains small in most of the underdoped (UD) regime, increases sharply to a peak slightly above optimal-doping (OPT), and then decreases in the overdoped (OD) regime. Our model predicts a similar doping evolution of the pseudogap and AN pair weight, blue and green lines in Fig.~\ref{fig:1}(a), respectively. The calculated pseudogap spectral weight is estimated as the change in spectral weight between the paramagnetic (PM) and AF phase (defined as the AF state at $T=0$ with SC order suppressed).  The AN pair weight is defined as the difference in AN spectral weight between the AF state and the zero temperature system with AF and SC order. The experimental data are defined by linear extrapolation of $T$-dependent trends to $T=0$, which may lead to some minor differences with our calculations. The sharp transition of spectral weight in the UD regime is a signature of TT1: when the AN electron pocket is lifted above $\epsilon_F$ with decreasing $x$, the AF order rapidly increases, but the SC pairing strength drops sharply when the pocket is too far from $\epsilon_F$ to contribute to pairing.  This TT occurs at $x_{LDA}=0.138$ (Due to complications when SC order is present, we define the crossing point in a model with AF order only, with SC order artificially set to zero.) marked as a black vertical dotted line in Fig.\ref{fig:1}(b), corresponding to the sudden increase in pair weight loss. At $x_{LDA}<0.12$, the electron pocket is $>15$~meV above $\epsilon_F$, too high in energy to induce any pairing.  At the same time, nodal pairing persists to considerably lower doping. 

Notably, TT1 has been found in Bi2201 from thermopower\textsuperscript{20} near $x_{USD}=0.13$ [green dashed line], and in STM\textsuperscript{15} [orange dot-dashed line] near $x_{USD}=0.1$, while a similar TT has been reported in Bi2212 near $x_{LDA}=0.1$\textsuperscript{21}. An analogous TT1 is also found in electron doped cuprates\textsuperscript{3,4}, where a hole pocket appears, centered around the antinode, near $-x=0.14-0.15$. While there is good agreement between the aforementioned estimates of TT1, $x_{TT1}\simeq 0.13$, the transition seen in ARPES\textsuperscript{5} appears to lie at a significantly higher doping, with sharp changes appearing in the overdoped regime.  This may indicate that ARPES is observing a different {\it TT}, possibly associated with a competing charge density wave (CDW) order not captured by the present model\textsuperscript{22,23,24,25,26,27}.  Note that in the ARPES\textsuperscript{5} data AN pair weight actually appears at a much lower doping, close to $x_{TT1}$, arrow in Fig.~\ref{fig:1}(b).

\begin{figure}[htp] 
\includegraphics[scale=0.9,keepaspectratio=true]{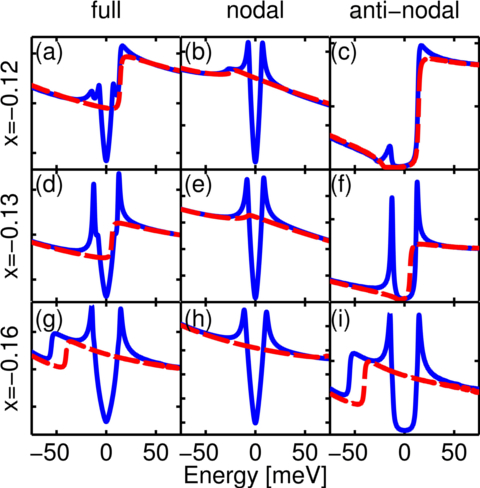}
\caption
{{\bf TT1 and FS-free SC in the DOS.} (a-i): DOS for the AF+SC system (blue curves) and the AF only system (red dashed curves).  The dopings shown are $x=0.12$ (a-c), $x=0.13$ (d-f), and $x=0.16$ (g-i). The first column (a,d,g) shows the full density of states. Here you can see the UMB crossing the fermi surface as doping is increased. It is also clear that there are two superconducting gaps. The outer gap in (a), $x=0.12$, is an example of FS-free superconductivity. The second column (b,e,h) is the density of states calculated only in the nodal region, 
as explained in the text, and contains only the inner gap seen in the full density of states. Similarly, the right-hand column (c,f,i) is the antinodal region, and encompasses only the antinodal electron pockets. 
As doping increases the electron pocket crosses the FS and we see the transition from a FS-free superconducting gap in (f) due to spectral weight being symmetrized about the FS to an ordinary gapped band in (i).
\label{fig:3}
}
\end{figure}

TT1 is reflected in many properties of the cuprates.  Fig.~\ref{fig:2}(a) shows that the self-consistent magnetization drops sharply across the transition as the electron pocket opens.  Fig.~\ref{fig:3} considers evolution of the SC gap and the density of states (DOS) in the vicinity of this TT. In the absence of superconductivity (red dashed curves), the bottom of the UMB shows up as a step increase in the DOS, starting just above $\epsilon_F$ at $x_{LDA}=0.138$, and moving well below $\epsilon_F$ at $x_{LDA}=0.16$, with the transition occuring at $x_{LDA}=0.138$. The doping dependence of the UMB results from the shrinking of $\Delta_{AF}$ with doping, Fig.~\ref{fig:2}(a). The middle column in Fig.~\ref{fig:3} shows the DOS calculated only in the region $(0<k_x<3/4\pi,0<k_y<3/4\pi)$, while the right column gives the DOS calculated in the remaining region of the first Brillouin zone. The middle column captures the spectral weight of the nodal states, and the right column that of the antinodal states. 

Evolution of the AF+SC state with doping in Fig. 3 (solid blue lines) is more complex. In the bottom row of Fig.~\ref{fig:3}, at $x_{LDA}=0.16$ the bottom of the UMB is well below $\epsilon_F$ above $T_c$. When SC turns on, the bottom of the UMB is shifted to lower energy, and the superconducting gap appears as a single feature. The SC gap is fairly symmetric, but with excess weight below $\epsilon_F$. At $x_{LDA}=0.13$ (middle row), the situation is completely changed. The bottom of the UMB is no longer seen clearly, but the SC gap now has two components, an inner gap and an outer gap, similar to what was found earlier in Bi2212\textsuperscript{21}. Note that the gap asymmetry has now reversed, with more weight above $\epsilon_F$. The interpretation of these features can be clarified with reference to the $x_{LDA}=0.12$ results in the top row of Fig. 3. Here the anisotropy is larger, indicating that the outer peak above $x_{LDA}=0.13$ is derived from the bottom of the UMB.  An analysis of the composition of the wave functions further confirms that at $x=0.12,$~0.13 there is a coherence peak in the AN region, even though the UMB would be entirely above the Fermi level in the absence of SC.  This is the essence of the phenomenon of {\it Fermi-surface-free superconductivity.} To understand the origin of the two gap features, we decompose the DOS into nodal parts, Figs.~\ref{fig:3}(b,e,h), and antinodal parts, Figs.~\ref{fig:3}(c,f,i).  The inner and outer gap can now be seen to arise from the nodal and antinodal regions, respectively. 
Note that at $x_{LDA}=0.12$ no part of the UMB crosses the FS.  However, the outer gap is produced by symmetrization of the UMB bottom around the FS in the SC state. Such a shift of spectral weight below $\epsilon_F$ leads to an enhanced stability of the SC state when the normal state DOS has an electron-hole asymmetry.

\begin{figure}[htp] 
\includegraphics[scale=0.9,keepaspectratio=true]{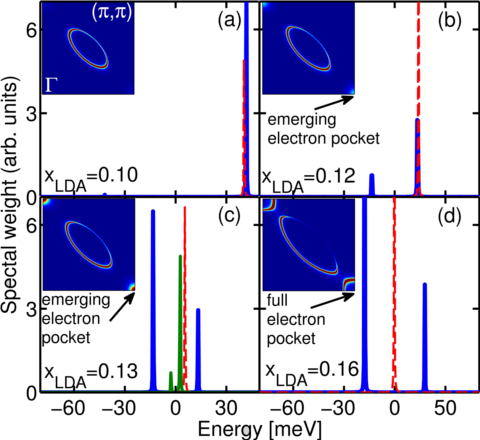}
\caption
{
{\bf TT1 and FS-free SC in Energy distribution curves.}  (a-d): Energy distribution curves (EDCs) of the AN point for dopings $x_{LDA}=0.10,0.12,0.13,$ and $0.16$ in (a,b,c,d) respectively. The insets show the momentum dependence of the spectral weight for an energy cut at the FS for a system with no SC gap. The red dashed curves show the EDC with no SC gap and the blue curves show the EDC with AF and SC gaps. The blue curves show the FS-free SC gap in frames (b) and (c), while (d) shows the situation after the topological transition. Here there is just a regular SC gap due to a band crossing the FS. This figure can be compared directly to Fig.~\ref{fig:3} which shows the transition from the perspective of DOS rather than a single momentum point. The green curve in (c) shows the EDC at $x=0.13$ for a linear $V$ as outlined in Fig.~\ref{fig:2}. Here we still observe FS-free SC below TT1 despite the reduced SC gap.    
\label{fig:4}
}
\end{figure}

Further insight is obtained from Figs.~\ref{fig:4}(a-d), which present energy distribution curves (EDCs) of the AN point where the spectral weight is given in  Fig.~\ref{fig:1}(a). When there is a well-defined electron pocket, as in Fig.~\ref{fig:4}(d) (inset), the AN EDC consists of a peak centered at the Fermi level (red dashed curves, which represent the EDCs for the non-superconducting state) however, for $x\le 0.138$, the EDC peak lies above $\epsilon_F$, so that the AN spectral weight is present only due to peak broadening. Thus, for $x_{LDA}=0.10$ the UMB has not yet crossed the FS and for $x_{LDA}=0.13$ the band is just beginning to cross. For $x_{LDA}=0.16$ the FS peak in the AF state is split when the SC turns on, and the spectral weight at the Fermi energy is greatly reduced.  

FS-free superconductivity is reflected in the EDCs of Fig. 4 before the onset of the TT through the presence of two peaks symmetric about the Fermi energy in the SC state (blue curves) at both $x_{LDA}=0.12$ and $x_{LDA}=0.13$, even though there are no bands crossing the Fermi energy. With increasing doping, as the UMB crosses the Fermi energy, the FS topology changes with the appearance of antinodal electron pockets centered around $(\pi,0)$. With reference to the spectral weight loss with and without superconductivity shown in Fig.~\ref{fig:1}(a), and the results of Figs.~\ref{fig:3} and \ref{fig:4}, we see that after the first TT around $x_{LDA}=0.12$, there is a sharp increase in the spectral weight loss. This weight loss is due to pair formation and is a direct result of the opening of antinodal electron pockets.

It is interesting to note that in order to reproduce the experimental dome in the low-doping regime, the interaction parameter $V$ in Fig.~\ref{fig:2}(b) would have to increase rapidly with underdoping below TT1. While a strong increase of the pairing potential near half-filling has been predicted\textsuperscript{28}, the dashed line in Fig.~\ref{fig:2}(b) indicates the effects of a more modest increase in $V$.  $T_c$ now decreases very rapidly below the first TT, Fig.~\ref{fig:2}(a), but there is still a range of FS-free superconductivity in Fig.~\ref{fig:4}(c).  To explain the lower part of the experimental SC dome in this scenario, we would have to postulate that the uniform AF+SC phase becomes unstable to nanoscale phase separation (NPS)\textsuperscript{29}, which is sensitive to impurities, and hence could lead to the observed low-energy spin-glass phase and to  the opening of an additional, nodal gap. Termination of this NPS at TT1 suggests that superconductivity stabilizes the associated $(\pi,\pi)$ magnetic order.

\section{Discussion}
We have shown that Fermi-surface-free superconductivity, as previously observed in the pnictides\textsuperscript{1,2}, can also occur in hole doped cuprates. This occurs near the doping at which topology of the Fermi surface changes as an electron pocket appears in the antinodal region, similar to the case of electron doped cuprates\textsuperscript{3,4}. The resulting spectral weight loss in the superconducting state is similar to that found in ARPES measurements\textsuperscript{5}. Our results provide evidence for the presence of two topological transitions under the superconducting dome in Bi2201, consistent with the picture of a trisected dome in Bi2212\textsuperscript{30}. 

\section{Methods}
Our analysis utilizes a mean field tight binding model with a Hubbard U along with SC and $(\pi,\pi)$-AF order as outlined in equations 2 and 3. The SC and AF order parameters are calculated self-consistently with equations 5 and 6. Within this framework, measurable quantities, such as spectral weight and density of states, can be calculated and analyzed accordingly. With this simple model we are able to describes the fundamentals of the TT and FS-free SC in hole doped cuprates.

The hopping parameters used here are based on photoemission experiment\textsuperscript{13}. We find that the resulting mean-field results are in good agreement with more accurate intermediate-coupling [quasiparticle-GW\textsuperscript{9}] results, when the following substitutions are made.  First, the experimental dispersion approximately matches the coherent part of the dressed bands, being renormalized by a factor $Z$ resulting in the relationship $\varepsilon_{exp}=Z\varepsilon_{LDA}$ where $\varepsilon_{exp}$ are the bands fit to experiment and $\varepsilon_{LDA}$ are bands fit to LDA calculations. When considering magnetic order $Z$ renormalizes the magnetic susceptibility $\chi_0$ and the Hubbard U. This leads to $U_{eff}\chi(\varepsilon_{exp})=U\chi_0(Z\varepsilon_{LDA})Z^2$, where $U_{eff}=Z^2U$ is the effective $U$ and $\chi(\varepsilon_{exp})=\chi_0(Z\varepsilon_{LDA})$ is the renormalized susceptibility. The resulting Stoner criterion is $U_{eff}\chi(\varepsilon_{exp})=1$. Therefore the $U$ considered in this paper is $U_{eff}$ which is what would be measured by experiments. 

\section{References}
\begin{enumerate}
\item Thirupathaiah, S. {\it et al.} Why $T_c$ of (CaFeAs)$_10$Pt$_{3.58}$As$_8$ is twice as high as (CaFe$_{0.95}$Pt$_{0.05}$As)$_{10}$Pt$_3$As$_8$. {\it Phys. Rev. B.} {\bf 88},140505(R) (2013).

\item Bang, Y. Shadow gap in the over-doped (Ba$_{1-x}$K$_x$)Fe$_2$As$_2$ compound. {\it arXiv}:1308.2413v1 (2013).

\item Kusko, C., Markiewicz, R. S., Lindroos, M. \& Bansil, A. Fermi surface evolution and collapse of the Mott pseudogap in Nd$_{2-x}$Ce$_x$CuO$_{4\pm\delta}$. {\it Phys. Rev. B.} {\bf 66}, 140513(R) (2002).

\item Das, T., Markiewicz, R. S., \& Bansil, A. Nodeless d-Wave Superconducting Pairing due to Residual Antiferromagnetism in Underdoped Pr$_{2-x}$Ce$_x$CuO$_{4-\delta}$. {\it Phys. Rev. Lett.} {\bf 98}, 197004 (2007).

\item Kondo, T. {\it et al.} Disentangling Cooper-pair formation above the transition temperature from the pseudogap state in the cuprates. {\it Nature Physics} {\bf 7} 21-25 (2011).

\item LeBlanc, J. P. F., Carbotte, J. P. \& Nicol, E. J. Signatures of Fermi surface reconstruction in Raman spectra of underdoped cuprates {\it Phys. Rev. B.} {\bf 81}, 064504 (2010). 

\item Das, T., Markiewicz, R. S., Bansil, A. \& Balatsky, A. V. Visualizing electron pockets in cuprate superconductors {\it Phys. Rev. B.} {\bf 85}, 224535 (2012).

\item Das, T., Markiewicz, R. S. \& Bansil, A. Nonmonotonic $d_{x^2-y^2}$ superconducting gap in electron-doped Pr$_{0.89}$LaCe$_{0.11}$CuO$_4$: Evidence of coexisting antiferromagnetism and superconductivity? {\it Phys. Rev. B.} {\bf 74},020506 (2006).

\item Das, T., Markiewicz, R. S. \& Bansil, A. Strong correlation effects and optical conductivity in electron-doped cuprates. {\it Europhys. Lett.} {\bf 96}, 27004 (2011).

\item Basak, S. {\it et al.} Origin of the high-energy kink in the photoemission spectrum of the high-temperature superconductor Bi$_2$Sr$_2$CaCu$_2$O$_8$. {\it Phys. Rev. B.} {\bf 80}, 214520 (2009).

\item Das, T., Markiewicz, R. S. \& Bansil, A. Optical model-solution to the competition between a pseudogap phase and a charge-transfer-gap phase in high-temperature cuprate superconductors. {\it Phys. Rev. B.} {\bf 81}, 174504 (2010).

\item Das, T., Markiewicz, R. S. \& Bansil, A. Reconstructing the bulk Fermi surface and superconducting gap properties from neutron scattering experiments. {\it Phys. Rev. B.} {\bf  85}, 064510 (2012).

\item He, R.-H. {\it et al.} From a Single-Band Metal to a High-Temperature Superconductor via Two Thermal Phase Transitions. {\it Science} {\bf 331}, 1579 (2011).

\item H{\"{u}}fner, S., Hossain, M. A., Damascelli, A. \& Sawatzky, G. A. Two gaps make a high-temperature superconductor? {\it Rep. Prog. Phys.} {\bf 71}, 062501 (2008).

\item He, Y. {\it et al.} Fermi Surface Pairing \& Coherence in a High Tc Superconductor. {\it arXiv}:1305.2778.

\item Piriou, A., Jenkins, N., Berthod, C., Maggio-Aprile, I. \& Fischer, \O. First direct observation of the Van Hove singularity in the tunneling spectra of cuprates. {\it Nat. Commun.} {\bf 2}, 221 (2011).

\item Presland, M., Tallon, J., Buckley, R., Liu, R. \& Flower, N. General trends in oxygen stoichiometry effects on $T_{c}$ in Bi and Tl superconductors. {\it Physica C} {\bf 176}, 95 (1991).

\item Markiewicz, R. S. Mode-coupling model of Mott gap collapse in the cuprates: Natural phase boundary for quantum critical points. {\it Phys. Rev. B.} {\bf 70} 174518 (2004).

\item Markiewicz, R. S. \& Bansil, A. Dispersion anomalies induced by the low-energy Plasmon in the cuprates. {\it Phys. Rev. B.} {bf 75} 020508 (2007).

\item Storey, J. G., Tallon, J. L. \& Williams, G. V. M. Electron pockets and pseudogap asymmetry observed in the thermopower of underdoped cuprates. {\it Europhys. Lett.} {\bf 102}, 37006 (2013).

\item Nieminen, J., Suominen, I., Das, T., Markiewicz, R. S. \& Bansil, A. Evidence of strong correlations at the van Hove singularity in the scanning tunneling spectra of superconducting Bi$_2$Sr$_2$CaCu$_2$O$_{8+\delta}$ single crystals. {\it Phys. Rev. B.} {\bf 85}, 214504 (2012).

\item Wu, T. {\it et al.} Magnetic-field-induced charge-stripe order in the high-temperature superconductor YBa$_2$Cu$_3$O$_y$. {\it Nature} {\bf 477}, 191 (2011).

\item Ghiringhelli, G. {\it et al.} Long-Range Incommensurate Charge Fluctuations in (Y,Nd)Ba$_2$Cu$_3$O$_{6+x}$. {\it Science} {\bf 337}, 821 (2012). 

\item Achkar, A. J. {\it et al.} Distinct Charge Orders in the Planes and Chains of Ortho-III-Ordered YBa$_2$Cu$_3$O$_{6+\delta}$ Superconductors Identified by Resonant Elastic X-ray Scattering. {\it Phys. Rev. Lett.} {\bf 109}, 167001 (2012).

\item Chang, J. {\it et al.} Direct observation of competition between superconductivity and charge density wave order in YBa$_2$Cu$_3$O$_{6.67}$. {\it Nature Physics} {\bf 8}, 871 (2012).

\item LeBoeuf, D. {\it et al.} Thermodynamic phase diagram of static charge order in underdoped YBa$_2$Cu$_3$O$_y$. {\it Nature Physics} {\bf 9},79 (2013). 

\item Blackburn, E. {\it et al.} X-Ray Diffraction Observations of a Charge-Density-Wave Order in Superconducting Ortho-II YBa$_2$Cu$_3$O$_{6.54}$ Single Crystals in Zero Magnetic Field. {\it Phys. Rev. Lett.} {\bf 110}, 137004 (2013). 

\item Maier, T. A., Jarrell, M. \& Scalapino, D. J. Pairing interaction in the two-dimensional Hubbard model studied with a dynamic cluster quantum Monte Carlo approximation. {\it Phys. Rev. B.} {\bf 74}, 094513 (2006).

\item Seibold, G., Markiewicz, R. S. \& Lorenzana, J. Spin canting as a result of the competition between stripes and spirals in cuprates. {\it Phys. Rev. B.} {\bf 83}, 205108 (2011).

\item Vishik, I. M. {\it et al.} Phase competition in trisected superconducting dome. {\it PNAS} {\bf 109}, 18332 (2012).
\end{enumerate}

\section{Acknowledgements}     
This work is supported by the US Department of Energy, Office of Science, Basic Energy Sciences grant number DE-FG02-07ER46352, and benefited from Northeastern University's Advanced Scientific Computation Center (ASCC), theory support at the Advanced Light Source, Berkeley and the allocation of supercomputer time at NERSC through grant number DE-AC02-05CH11231.

\section{Author contributions}
P.M., H.H., R.S.M., and A.B. contributed to the research reported in this study and the writing of the manuscript.

\section{Additional information}
The authors declare no competing financial interests. 

\section{Figures}

\end{document}